\documentclass[10pt]{iopart}

\usepackage{graphicx}  
\usepackage{dcolumn}
\usepackage{bm}

\begin{document}

\title[Effect of antisites in CrAs and related compounds]{Role of the presence of
transition-metal atoms at the antisites in CrAs, CrSe and VAs zinc-blende compounds}

\author{K \"Ozdo\~gan\dag, I Galanakis\ddag, B Akta\c s\dag, and E \c Sa\c s\i
o\~glu\S}

\address{\dag\ Department of Physics, Gebze Institute of Technology,
Gebze, 41400, Kocaeli, Turkey }

\address{\ddag\ Department of Materials Science, School of Natural
  Sciences, University of Patras,  GR-26504 Patra, Greece}

\address{\S\ Institut f\"ur Festk\"orperforschung, Forschungszentrum
J\"ulich, D-52425 J\"ulich, Germany and  Fatih University, Physics
Department, 34500, B\" uy\" uk\c cekmece,  \.{I}stanbul, Turkey}

\ead{kozdogan@gyte.edu.tr,galanakis@upatras.gr,e.sasioglu@fz-juelich.de}

\begin{abstract}
In a recent publication [Galanakis I et al 2006 \PR B \textbf{74}
140408(R)] we have shown that in the case of CrAs and related
transition-metal chalcogenides and pnictides, crystallizing in the
zinc-blende structure, the excess of the transition-metal atoms
leads to half-metallic ferrimagnetism. The latter property is
crucial for spintronic applications with respect to ferromagnets
due to the lower stray fields created by these materials. We
extend this study to cover the case where the transition-metal
atoms sitting at antisites are not identical to the ones in the
perfect sites. In Cr-based compounds, the creation of Mn antisites
keeps the half-metallic ferrimagnetic character produced also by
the Cr antisites. In the case of VAs, Cr and Mn antisites keep the
half-metallic character of VAs (contrary to V antisites) due to
the larger exchange-splitting exhibited by these atoms.
\end{abstract}


\section{Introduction}\label{sec1}

The rapid emergence of the field of spintronics (also known as
magnetoelectronics \cite{Zutic}) brought to the center of
scientific research the so-called half-metallic ferromagnets (like
Heusler alloys \cite{deGroot,GalanakisHalf,GalanakisFull} or some
oxides \cite{Soulen}).  These compounds present metallic behavior
for one spin-band while they are semiconducting or insulators for
the other spin-band, resulting to perfect spin-polarization, at
least for the bulk, at the Fermi level. For realistic application
ferromagnets create large stray fields and thus lead to
considerable undesirable energy losses. Thus, to this respect the
case of half-metallic ferrimagnets is more interesting. There are
several ways to create a half-metallic ferrimagnet, either by
doping a semiconductor like FeVSb \cite{Leuken} or due to the
simultaneous presence of Mn and another transition-metal atom in
the case of alloys with small total moment due to the small number
of valence electrons (e.g. FeMnSb \cite{GalanakisHalf} or
Mn$_2$VAl \cite{GalanakisFull,Kemal}). Recently Akai and Ogura
have proposed another route to fully-compensated half-metallic
ferrimagnetism based on the doping of diluted magnetic
semiconductors \cite{Akai}.

Except Heusler and oxides, also transition-metal chalcogenides
like CrAs and pnictides like CrSe are known to present
half-metallic ferromagnetism when they crystallize in the
metastable zinc-blende structure. The first experimental evidence
was provided in the case of CrAs thin-films  by the group of
Akinaga in 2000 \cite{Akinaga2000} and many more experiments have
confirmed these results \cite{experiments}. Experiments agree with
prediction of ab-initio calculations performed by several groups
\cite{MavropoulosZB,GalaZB,calculations,Shirai}. In the case of
the half-metallic ferromagnets like CrAs or CrSe, the gap in the
minority-spin band arises from the hybridization between the
p-states of the $sp$ atom and the triple-degenerated $t_{2g}$
states  of the transition-metal and as a result the total
spin-moment, $M_t$, follows the Slater-Pauling (SP) behavior being
equal in $\mu_B$ to $Z_t-8$ where $Z_t$ the total number of
valence electrons in the unit cell \cite{MavropoulosZB}. Recently
theoretical works have appeared attacking also some crucial
aspects of these alloys like  the exchange bias in
ferro-/antiferromagnetic interfaces \cite{Nakamura2006}, the
stability of the zinc-blende structure \cite{Xie2003}, the
dynamical correlations \cite{Chioncel2006}, the interfaces with
semiconductors \cite{Interfaces}, the exchange interaction
\cite{Sasioglu-Gala} and  the temperature effects
\cite{MavropoulosTemp}.

In a recent publication \cite{Rapid}, we have investigated the
properties of several compounds (CrAs, CrSb, CrSe, CrTe, VAs and
MnAs) when we create an excess of the transition metal atoms;
transition metal atoms occupy also antisites occupied in the
perfect compound by the $sp$ atom. To perform this study we used
the full--potential nonorthogonal local--orbital minimum--basis
band structure scheme (FPLO) and we simulated disorder via the
coherent potential approximation \cite{koepernik}; we also used
the theoretical equilibrium lattice constants for which the
perfect compounds are half-metallic ferromagnets \cite{lattices}.
The most interesting case was the Cr-based compounds. The
Cr-impurities in the case of antisites at the sublattice occupied
by the $sp$ atoms couple antiferromagnetically to the existing Cr
atoms at the ideal sites and destroy ferromagnetism. But these
compounds stay half-metallic for large concentration of antisites
exhibiting half-metallic ferrimagnetism. Also in
V$_{1+x}$As$_{1-x}$ compound the V atoms at the antisites couple
antiferromagentically to the V atoms at the perfect sites but the
exchange splitting of the V impurities is small and the Fermi
level cross the states of these atoms. In this manuscript we
expand this study to cover also the case where at the antisites we
find transition metal atoms of different chemical type than the
perfect sites. More precisely we studied the case of Mn impurites
in Cr[As$_1-x$Mn$_x$] and  Cr[Se$_1-x$Mn$_x$] and the case of Cr
and Mn impurities in  V[As$_1-x$Cr$_x$]  and V[As$_1-x$Mn$_x$]
compounds. We find that in the case of Cr-compounds the
half-metallic ferrimagnetic behavior is again present, and the
same is also true for the V[As$_1-x$(Cr or Mn)$_x$] compounds
since Cr and Mn atoms couple antiferomagnetically to the V atoms
at the perfect site and have a larger exchange splitting than the
V impurities in V$_{1+x}$As$_{1-x}$.

\begin{figure}
\begin{center}
\includegraphics[scale=0.5]{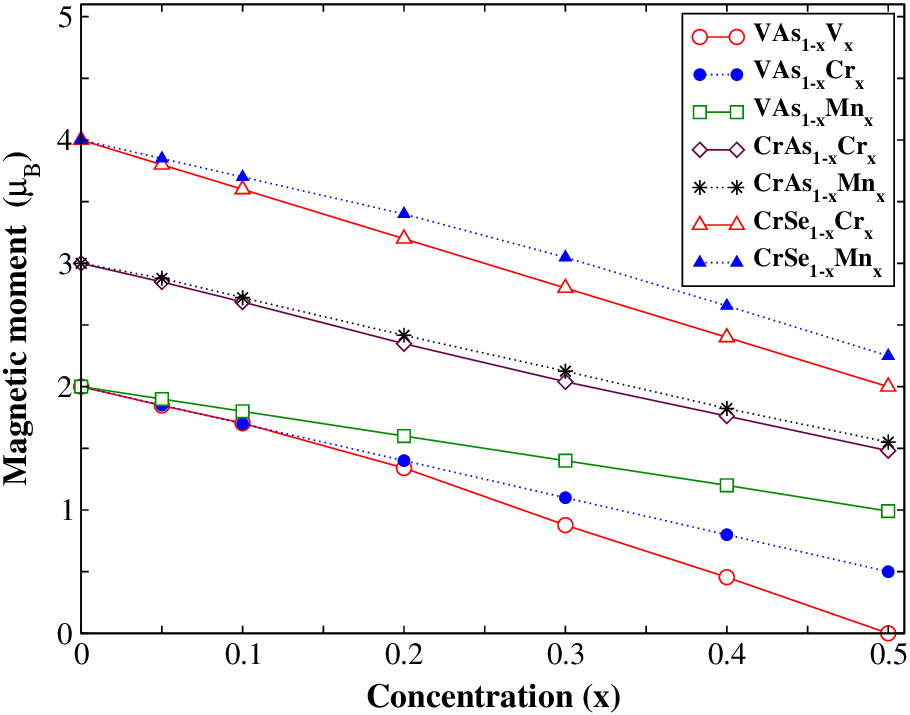}
\end{center} \caption{(Color online) Total spin magnetic moment as
a function of the concentration $x$ for the studied
T[Z$_{1-x}$T'$_{x}]$ compounds.\label{fig1}}
\end{figure}

\section{Total magnetic moments}\label{sec2}

Before going in details of the density of states (DOS) or atomic
spin moments we will present our results on the total spin
moments. In figure \ref{fig1} we have included the different cases
under study going up to x=0.5 where half of the $sp$ atoms have
been substituted by transition metal atoms. Of course such a case
is unlikely to be stable in the zinc-blende structure but we have
gone so far to make physics more transparent.The perfect CrAs
compound has a total spin moment in the unit cell of 3 $\mu_B$
since there are 11 valence electrons, CrSe with one valence
electron more in the unit cell has a spin moment of 4 $\mu_B$ and
VAs with 10 valence electrons a spin moment of 2 $\mu_B$. In all
cases the total spin moment decreases with the concentration, a
clear sign of the antiferromagnetic coupling between the
transition metal atoms at different sites. In the case of CrAs,
substituting Cr or Mn for As results to identical behavior (solid
black lines with open diamonds and stars, respectively in the
figure) and the total spin moment is almost independent of the
chemical type of the transition-metal atom at the antisite. This
is also true to a lesser extent for CrSe (red open triangles for
Cr[Se$_{1-x}$Cr$_x$] and blue triangles for Cr[Se$_{1-x}$Mn$_x$]).
For the same concentration x the compound with Mn has a slightly
larger total spin moment and as the concentration increases this
difference also increases.

In the case of VAs the differences are more pronounced. When we
substitute V for As, V[As$_{1-x}$V$_x$], the V atoms at the ideal
and the antisites are antiferromagnetically coupled. As the
concentration increases the absolute value of the V spin moments
decreases and for x=0.5 the total spin moment is almost zero and
the V atoms are almost non-magnetic. Cr and Mn impurities, on the
other hand, lead to larger total spin moment and the resulting
compounds are ferrimagnets with strong magnetic elements. Overall
the V[As$_{1-x}$Mn$_x$] compounds have a larger spin moment than
the V[As$_{1-x}$Cr$_x$] ones. We should note that although the
behavior of the total spin moments reveals the existence of
ferrimagnetism it does not give any information about the
half-metallic character and we should look how the DOS changes
with the concentration $x$ for every case to be sure if the
half-metallicity is preserved or lost.

\begin{table}
\caption{Total and atom-resolved spin magnetic moments for the
Cr[Z$_{1-x}$Y$_{x}$] compounds where Z= As or Se and Y= Cr or Mn.
As "imp" we denote the transition-metal atoms which are located at
antisite positions. Note that the atomic moments have been scaled
to one atom.} \begin{indented}
 \item[]
 \begin{tabular}{l|cccc|cccc} \hline \hline
 & \multicolumn{4}{c|}{Cr[As$_{1-x}$Cr$_{x}$]} & \multicolumn{4}{c}{Cr[As$_{1-x}$Mn$_{x}$]}\\
 $x$  & Total  &  Cr  & As & Cr-imp & Total & Cr & As & Mn-imp\\
 \hline
  0   & 3.00  & 3.41 &  -0.41 & --    & 3.00 & 3.41 & -0.41 & --    \\
 0.05 & 2.85  & 3.37 &  -0.40 & -2.77 & 2.88 & 3.38 & -0.40 & -2.49 \\
 0.1  & 2.69  & 3.33 &  -0.40 & -2.81 & 2.72 & 3.33 & -0.39 & -2.57 \\
 0.2  & 2.35  & 3.24 &  -0.40 & -2.88 & 2.42 & 3.25 & -0.38 & -2.66 \\
 0.4  & 1.76  & 3.19 &  -0.39 & -3.00 & 1.82 & 3.19 & -0.36 & -2.87 \\ \hline \hline
 & \multicolumn{4}{c|}{Cr[Se$_{1-x}$Cr$_{x}$]} & \multicolumn{4}{c}{Cr[Se$_{1-x}$Mn$_{x}$]}\\
 $x$  & Total &  Cr  & Se & Cr-imp & Total & Cr & Se & Mn-imp\\
 \hline
 0    &  4.00 & 4.20 & -0.20 & --    & 4.00 & 4.20 & -0.20 & --  \\
0.05  &  3.80 & 4.12 & -0.17 & -3.08 & 3.85 & 4.15 & -0.17 & -2.92  \\
0.1   &  3.60 & 4.06 & -0.17 & -3.08 & 3.70 & 4.14 & -0.16 & -2.92 \\
0.2   &  3.20 & 3.96 & -0.17 & -3.08 & 3.40 & 4.11 & -0.16 & -2.93\\
0.4   &  2.40 & 3.75 & -0.18 & -3.09 & 2.66 & 3.98 & -0.16 & -3.09\\
\hline \hline
\end{tabular}
\end{indented}
\label{table1}
\end{table}

\section{The Cr-based compounds}\label{sec3}

\begin{figure}
\begin{center}
\includegraphics[scale=0.6]{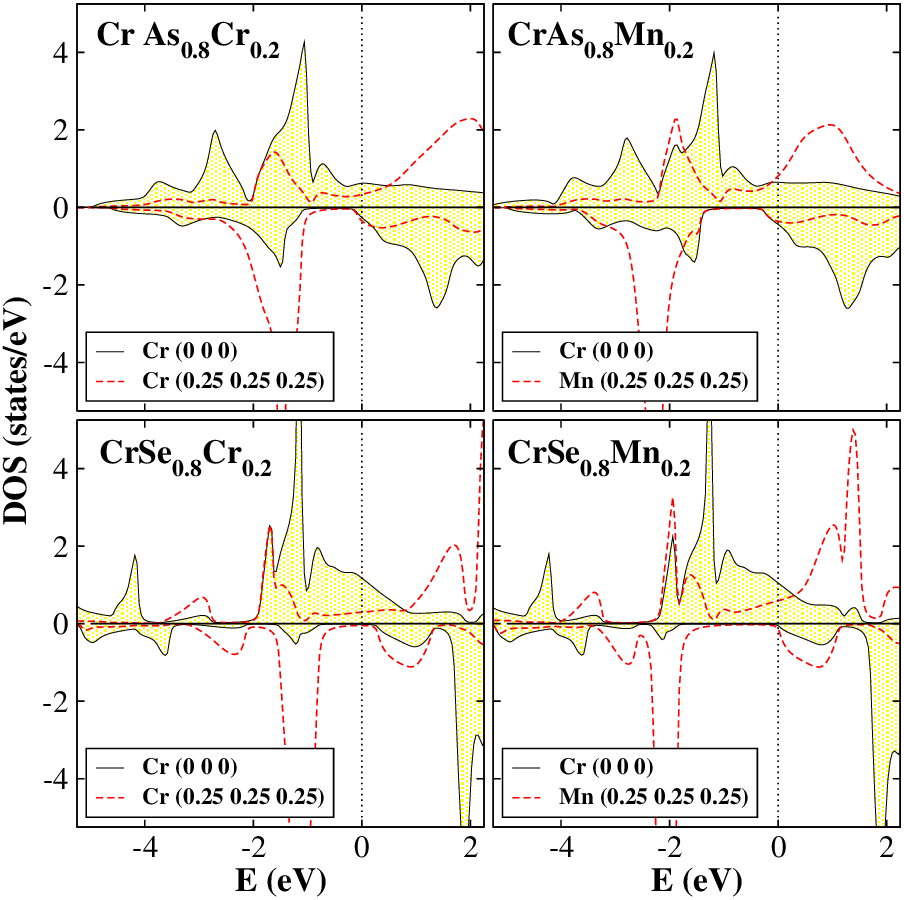}
\end{center} \caption{(Color online) Atom-resolved DOS for the Cr atom at the ideal (0 0 0) site and for the
transition-metal impurity at the $(\frac{1}{4}\: \frac{1}{4}\:
\frac{1}{4})$ antisite for the Cr[Z$_{0.8}$Y$_{0.2}$] compounds
where Z= As or Se and Y= Cr or Mn. Positive values of DOS
correspond to spin-up electrons and negative values to spin-down
electrons. The energy scale is such that the zero energy
corresponds to the energy of the Fermi level. Notice that the
majority electrons correspond to spin-up electrons for the Ct at
the ideal sites and to spin-down electrons for the Cr or Mn
impurities at the antisites.  The vice versa is true for the
minority-spin electrons. \label{fig2}}
\end{figure}

The first family of compounds under study is the CrAs and CrSe
alloys. We extended the study presented in reference \cite{Rapid}
to cover also the case when we substitute Mn for As or Se. CrSb is
isoelectronic to CrAs (same number of valence electrons) and CrTe
is isoelectronic to CrSe, thus they are expected to present
similar behavior and we have not included them in our study. In
table \ref{table1} we present the total and atom-resolved spin
magnetic moments for the Cr[As$_{1-x}$Cr$_{x}$] and
Cr[As$_{1-x}$Mn$_{x}$] compounds in the upper panel and the
Cr[Se$_{1-x}$Cr$_{x}$] and Cr[Se$_{1-x}$Mn$_{x}$] compounds in the
lower panel. In figure \ref{fig2} we present the atom-resolved DOS
for all four cases and for x=0.2.

In CrAs the Fermi level is close the upper-energy edge of the gap
and in CrSe somewhat deeper in the gap. Thus half-metallicity is
easier to be kept for the Se-compound. Overall the Cr atoms at the
perfect $(0\: 0\: 0)$ site show similar DOS for all cases and the
spin moment of these Cr atoms, as can be seen in table
\ref{table1}, present small variations with the impurities
concentration at the antisites (the differences are somewhat
larger for the Se compounds). Each Cr atom at the perfect site has
four As or Se  nearest-neighbors. As we substitute Cr or Mn for As
or Se, the immediate environment of the Cr atoms at the perfect
sites changes and instead of hybridizing with the p-states of As
or Se now they hybridize with the $t_{2g}$ states of Cr or Mn
impurities leading to a small decrease of the spin moment of these
Cr atoms at the ideal sites.

Both Cr and Mn impurities have all their five majority d-states
(spin-down states with respect to the Cr atoms at the perfect
sites) occupied as well as the minority (spin-up)
double-degenerated $e_g$ states occupied. The spin moment at the
impurity is around -3 $\mu_B$ being the total number of the
uncompensated spins. This is clearly seen in figure \ref{fig2}
where we present with red dashed lines the DOS of the impurity
atoms at the antisites. For majority spins (negative values of
DOS) there is one large peak occupied containing all five
d-states. For the minority spins (positive values of DOS) there is
a much smaller peak occupied containing the double-degenerated
$e_g$ electrons. The $e_g$ electrons are well localized since they
do not hybridize due to symmetry reasons neither with the $t_{2g}$
electrons of the transition metal atoms nor with the $p$ electrons
of As or Se and thus they are occupied for both spin directions.
On the other hand the triple-degenerated $t_{2g}$ states present
strong hybridization with their environment and a considerable
large exchange splitting. As a result the majority $t_{2g}$ states
are completely occupied and the minority $t_{2g}$ states are empty
leading to the spin moment of around -3 $\mu_B$. Mn presents a
larger exchange-splitting than Cr and thus also in its case the
spin-moment can not surpass the value of -3. The small deviations
from this ideal value in table \ref{table1} come from the
hybridization of the Cr or Mn impurities with the d-states of the
Cr atoms at the perfect sites.

As we have already mentioned the Cr atoms at the ideal $(0\: 0\:
0)$ sites do not change their DOS with the concentration of the Cr
or Mn impurities and they do not affect the gap (neither its
location or its width). Also Mn and Cr impurities present a large
exchange splitting and there is a considerable gap between the
occupied majority (spin-down) d-states and the unoccupied
antibonding states. The width of this gap is much larger for the
Mn impurities (this is clear if we look at the CrSe case) where
the exchange splitting between the majority (spin-down) occupied
d-states and the minority (spin-up) unoccupied $t_{2g}$-states is
larger than for the Cr atoms. All these phenomena result in the
persistence of the gap and thus of the half-metallicity upon
doping with either Cr or Mn for As or Se. Also as we mentioned
already the impurities are antiferromagentically coupled to the Cr
atoms at the ideal sites leading to the desirable half-metallic
ferrimagnetism.

Before proceeding to the V-based compounds we have to shortly also
discuss the behavior of the sp atoms. Both As and Se have their
minority p-states completely occupied while for the majority
p-states some of the weight is over the Fermi level leading to
small negative spin moments as seen in table \ref{table1}. The
origin of the antiferrimagnetic coupling between the
transition-metal atom and the sp-atom is discussed in detail in
reference \cite{GalaZB}. Thus the change in the concentration of
As or Se only slightly affects the properties of these atoms and
the small change in their spin moments arises mainly due to the
change in the spin moments of the Cr atoms at the ideal site
through the hybridization between the Cr d states and the As or Se
p states.

\begin{figure}
\begin{center}
\includegraphics[scale=0.55]{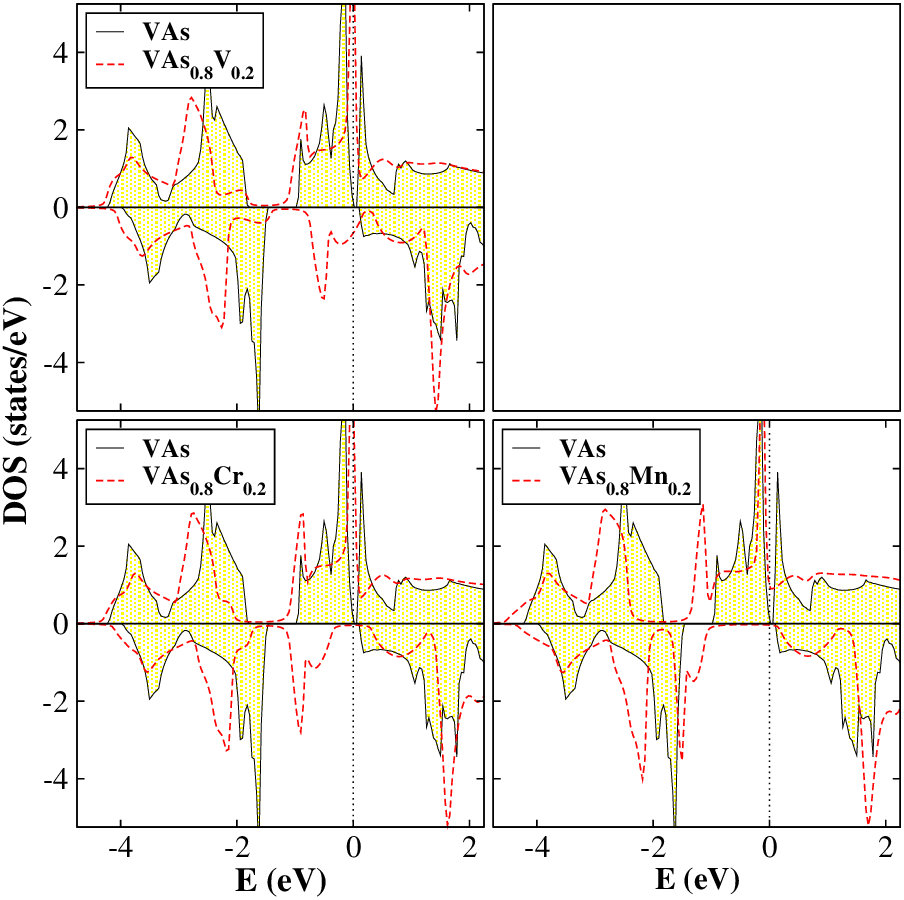}
\end{center} \caption{(Color online) Total DOS for the V[As$_{0.8}$V$_{0.2}$] (upper panel),
V[As$_{0.8}$Cr$_{0.2}$] (lower left panel) and
V[As$_{0.8}$Mn$_{0.2}$] (lower right panel) compounds with respect
to the perfect VAs (shaded area). \label{fig3}}
\end{figure}

\section{The V-based compounds}\label{sec4}

\begin{figure}
\begin{center}
\includegraphics[scale=0.6]{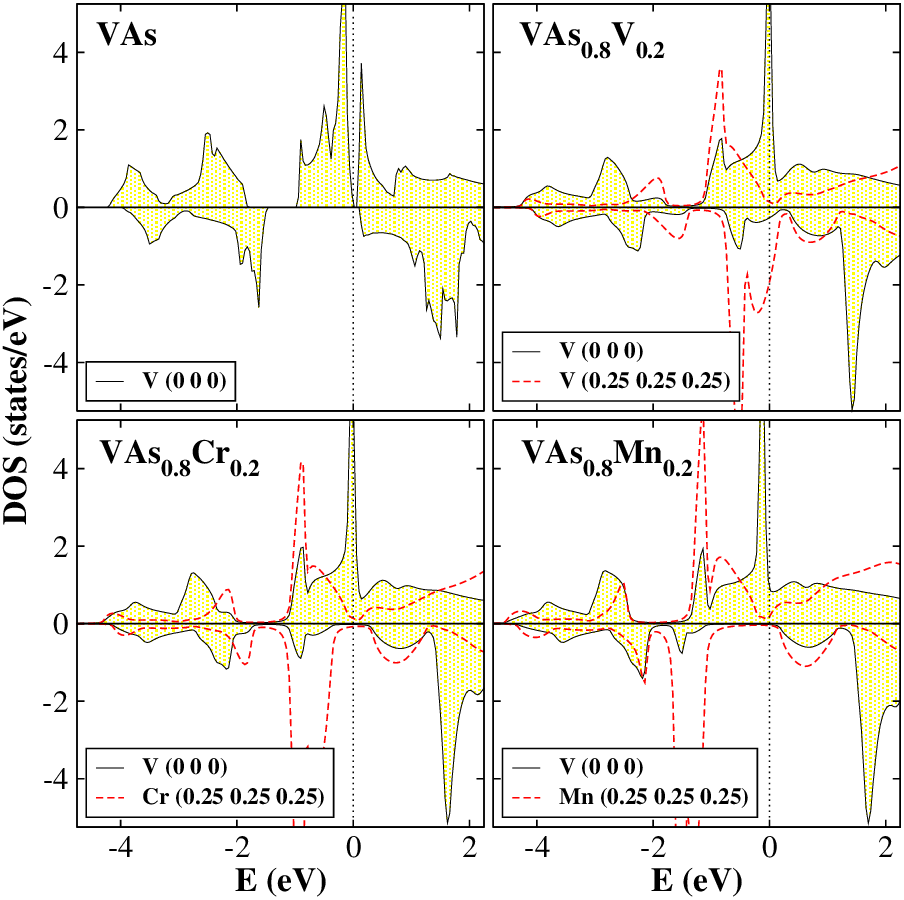}
\end{center} \caption{(Color online) Atom-resolved DOS for the V atom at the ideal (0 0 0) site and for the
transition-metal impurity at the $(\frac{1}{4}\: \frac{1}{4}\:
\frac{1}{4})$ antisite for the V[As$_{0.8}$Y$_{0.2}$] compounds
where Y= V, Cr or Mn.  \label{fig4}}
\end{figure}

The second case under study is the presence of Cr and Mn
impurities in VAs. In reference \cite{Rapid}, we have shown that
the excess of V, V[As$_{1-x}$V$_x$] family, destroys the
half-metallic character of the ideal VAs alloy. In figure
\ref{fig3} we have gathered the total DOS (red dashed line) for
the V[As$_{0.8}$Y$_0.2$] with Y= V, Cr or Mn compared to the total
DOS for the ideal VAs alloy (shaded area). It is obvious that Cr
and Mn impurities due to the larger exchange splitting, which they
present with respect to the V impurities, keep the half-metallic
character although especially for Cr the width of the gap is
severely reduced; for Mn the occupied spin-down states are pushed
very low in energy and the width of the gap is almost unchanged.
We will discuss in detail this phenomenon in the following
paragraph. For now let's concentrate on the behavior around the
Fermi level of the majority DOS. In perfect VAs the Fermi level
falls within a very sallow majority gap so that only the majority
$e_g$ states are occupied leading to a total spin moment of
exactly 2 $\mu_B$ \cite{GalaZB}. When we substitute V, Cr or Mn
for As, even at low concentration of the transition-metal
impurities, the large majority peak of V just below the Fermi
level moves slightly higher in energy and now the Fermi level
falls within this peak. In figure \ref{fig4} where we present the
atom-resolved DOS, it is clear that this peak comes exclusively
from the V atoms at the ideal $(0\: 0\: 0)$ sites and is
irrelevant to the impurities.

As mentioned above in figure \ref{fig4} we have drawn the
atom-resolved DOS for the transition metal atoms in the case of
the perfect VAs and the V[As$_{0.8}$Y$_0.2$] with Y= V, Cr or Mn
alloys(V atoms at the ideal $(0\: 0\: 0)$ sites with the shaded
region and V, Cr or Mn impurities at the $(0.25\: 0.25\: 0.25)$
antisites with the dashed red lines). In table \ref{table2} we
have gathered the total and spin magnetic moments for the
V[As$_{1-x}$Cr$_{x}$] and V[As$_{1-x}$Mn$_{x}$] compounds as a
function of the concentration of the impurities. We will start our
discussion from the V atoms at the ideal sites. Due to the change
in their environment $e_g$ and $t_{2g}$ majority states are not
any more separated by a tiny gap but they slightly overlap so that
that the Fermi level is not in a majority tiny gap any more.
Comparing the values for V[As$_{1-x}$V$_{x}$] in table 1 of
reference \cite{Rapid} and the values for V[As$_{1-x}$Cr$_{x}$]
and V[As$_{1-x}$Mn$_{x}$] in table \ref{table2} of the present
manuscript, we can identify a global trend in the behavior of the
V at the perfect site. As we move from the heavier element Mn to
the lighter V, the increase of the impurities concentration pushes
the Fermi level deeper in the large peak located around the Fermi
level resulting to smaller V spin magnetic moments. To make it
more clear we consider the case for x=0.2. In the case of V
impurities the V at the ideal site has a spin moment of 1.7
$\mu_B$, for Cr impurities its spin moment is 1.98 $\mu_B$ while
for  V[As$_{0.8}$Mn$_{0.2}$] the spin moment of the V at the ideal
site is 2.21 $\mu_B$ very close to the value for the perfect VAs
compound of 2.27 $\mu_B$. Due to this larger drop in the spin
moment of the V atoms at the perfect site for the lighter element,
we observe the behavior of the total spin moment in figure
\ref{fig1}. For V impurities the total spin moment is almost zero
(in reality slightly negative) for x=0.5 and for the case of Cr
impurities the total spin moment is 0.5 $\mu_B$ for z=0.5. On the
other hand for Mn impurities the spin moment of V atoms remains
almost constant as we can see in table \ref{table2} and the drop
in the total moment comes exclusively from the increase in the
concentration of Mn being ~1 $\mu_B$ for x=0.5 double the one for
the Cr impurities case.

\begin{table}
\caption{Total and atom-resolved spin magnetic moments for the
V[As$_{1-x}$Y$_{x}$] compounds where Y= Cr or Mn. As "imp" we
denote the transition-metal atoms which are located at antisite
positions. Note that the atomic moments have been scaled to one
atom.} \begin{indented}
 \item[]
 \begin{tabular}{l|cccc|cccc} \hline \hline
 & \multicolumn{4}{c|}{V[As$_{1-x}$Cr$_{x}$]} & \multicolumn{4}{c}{V[As$_{1-x}$Mn$_{x}$]}\\
 $x$  & Total  &  V  & As & Cr-imp & Total & V & As & Mn-imp\\
 \hline
  0   & 2.00  & 2.27 &  -0.27 & --    & 2.00 & 2.27 & -0.27 & --    \\
 0.05 & 1.85  & 2.19 &  -0.25 & -2.07 & 1.90 & 2.25 & -0.26 & -2.02 \\
 0.1  & 1.70  & 2.13 &  -0.25 & -2.05 & 1.80 & 2.23 & -0.25 & -2.03 \\
 0.2  & 1.40  & 1.98 &  -0.25 & -1.99 & 1.60 & 2.21 & -0.25 & -2.05 \\
 0.3  & 1.10  & 1.82 & -0.21  & -1.91 & 1.40 & 2.19 & -0.25 &
 -2.07 \\
 0.4  & 0.80  & 1.66 & -0.19  & -1.86 & 1.20 & 2.18 & -0.24 &-2.10\\
 0.5  & 0.50  & 1.56 & -0.18  & -1.93 & 0.99 & 2.19 & -0.24 &
 -2.15
 \\ \hline \hline
\end{tabular}
\end{indented}
 \label{table2}
\end{table}

We still have not discussed the behavior of the impurity atoms in
the case of VAs. From the table \ref{table2} it is obvious that,
contrary to Cr-based compounds, the impurity Cr or Mn atoms have a
spin magnetic moment of around -2 $\mu_B$, while the V impurity
atoms (see table 1 in reference \cite{Rapid}) have a spin moment
of around -1 $\mu_B$. To understand this behavior we have to look
at the atom-resolved DOS in figure \ref{fig4}. The exchange
splitting of the V impurities at the antisites is very small and
the majority (spin-down) and minority (spin-up) double degenerated
$e_g$-states are completely occupied (peaks with the higher
intensity located around -1 eV). Also one of the minority and two
of the majority triple-degenerated $t_{2g}$ states are occupied
resulting in a spin moment of around -1 $\mu_B$ and the Fermi
level is crossing the spin-down band of the $t_{2g}$ states. For
Cr and Mn impurities the exchange splitting is larger and all
three majority (spin-down) $t_{2g}$-states are occupied while the
picture for the minority states does not change and consecutively
the spin moment of the impurity is around -2 $\mu_B$. Also this
results to a spin-down gap separating the fully occupied d-states
from the antibonding states lying higher in energy and thus to
half-metallic ferrimagnetism. The exchange splitting is larger for
Mn than Cr and the spin-down gap is larger in the former case. As
we change the concentration of the impurity atoms their spin
moment only slightly changes due to the stronger hybridization
with the V atoms at the ideal sites. Finally we should note that
as we decrease the number of valence electrons and pass from CrSe
to CrAs and then to VAs the ionicity of the compounds becomes
smaller and so do the gaps created by the impurity atoms (smaller
exchanged splitting for the same impurity atom when it is found in
environment with smaller ionicity).

\section{Conclusion}\label{sec5}

We have complemented our study presented in a recent publication
 [Galanakis I et al 2006 \textit{Phys. Rev B} \textbf{74} 140408(R)] where we have
shown that in the case of CrAs and related transition-metal
chalcogenides and pnictides, crystallizing in the zinc-blende
structure, the excess of the transition-metal atoms leads to
half-metallic ferrimagnetism. In this contribution we show that
there may be additional advantages when the excess of the
transition-metal atom is not of the same chemical type as the one
at the perfect lattice sites.  In CrAs and CrSe, the creation of
Mn antisites keeps the half-metallic character and the Mn
impurities are antiferromagnetically coupled to the Cr ones
leading to half-metallic ferrimagnetism. The larger exchange
splitting of the Mn atoms with respect to the Cr ones makes the
half-metallic ferrimagnetism even more robust since the gap-width
is larger with respect to the case of Cr antisites. Even in VAs,
which looses its half-metallic character upon creation of V
antisites, the appearance of Cr or Mn antisites is accompanied
with the appearance of half-metallic ferrimagnetism. These results
may suggest a new way to achieve stable half-metallic
ferrimagnets, which is crucial for spintronic applications with
respect to half-metallic ferromagnets due to the lower stray
fields created by these materials.

\ack Authors  acknowledge  the computer support of the ``Leibniz
Institute for Solid State and Materials Research Dresden''.


\end{document}